\documentclass[oneside,english]{amsart}
\usepackage[T1]{fontenc}
\usepackage[latin9]{inputenc}
\usepackage{babel}
\usepackage{mathtools}
\usepackage{amstext}
\usepackage{amsthm}
\usepackage{amssymb}
\usepackage[pdfusetitle,
 bookmarks=true,bookmarksnumbered=false,bookmarksopen=false,
 breaklinks=false,pdfborder={0 0 1},backref=false,colorlinks=false]
 {hyperref}
\usepackage{stmaryrd}

\makeatletter
\numberwithin{equation}{section}
\numberwithin{figure}{section}
\theoremstyle{plain}

\theoremstyle{definition}

\theoremstyle{plain}

\theoremstyle{remark}
\newtheorem*{rem*}{\protect\remarkname}
\theoremstyle{plain}

\theoremstyle{definition}
\newtheorem*{defn*}{\protect\definitionname}
\theoremstyle{plain}

\theoremstyle{remark}

\newcommand{\lean}[1]{{\ttfamily\small #1}}
\makeatother

\usepackage{listings}
\usepackage{booktabs}
\usepackage{array}
\providecommand{\definitionname}{Definition}
\providecommand{\corollaryname}{Corollary}
\providecommand{\lemmaname}{Lemma}
\providecommand{\propositionname}{Proposition}
\providecommand{\remarkname}{Remark}
\providecommand{\theoremname}{Theorem}

\begin{document}
\title[CrossCheck II: Certified Cost Bounds for Real-Time C and C++]{CrossCheck II: A Compiler Plugin for Certified Cost Bounds of Real-Time C and C++ Code, and its Evaluation}
\author{Luca Dall'Olio}
\thanks{Alter Ego Engineering, 5 rue du Docteur Lequeux, 92330 Sceaux, France. Email: \texttt{ld@aego.ai}. ORCID: \href{https://orcid.org/0009-0006-8839-5909}{0009-0006-8839-5909}. This preprint: \href{https://doi.org/10.5281/zenodo.24444646}{doi:10.5281/zenodo.24444646}.}
\begin{abstract}
We describe a plugin for the Clang compiler that bounds the cost of C and C++ functions intended for real-time use, and rejects at compile time the functions that allocate, throw, call through pointers, or contain a loop or a recursion for which no bound is found. Loops are bounded by affine ranking potentials synthesized with an SMT solver through Farkas' Lemma, or by logarithmic ones, recursion by depth and size potentials, and callees are analyzed in the calling context in which they are used. For functions in a linear fragment, the plugin emits a cost certificate that a checker, compiled from a Lean~4 definition proved sound in a companion paper, validates with integer arithmetic only; the bound then depends neither on the analyzer nor on the solver. On the 29 kernels of the TACLeBench collection, without annotations, the plugin bounds 142 of 220 loops in the calling contexts of the programs, 131 of them exactly as the benchmarks' flow facts, and 161 of the 165 bounded functions receive a certificate that the checker accepts, 152 of them also validated against an independent description of the source. On the CMSIS-DSP library for Cortex-M (786 functions), with assumed input ranges, it bounds 616 functions. We compare the loop bounds with loop counts observed by running the kernels and with five existing analyzers (LLVM's ScalarEvolution, KoAT2, CoFloCo, Infer and oRange), and the plugin with KoAT2 and CoFloCo on the C programs of the Termination and Complexity Competition: without annotations, 147 of 553 programs receive a certified bound that holds for all inputs, where the two tools bound 288 and 264. Loop bounds, exported annotations and the checker program are also tested on random programs and on mutated certificates.
\end{abstract}

\maketitle

\section{Introduction}

Hard real-time software must finish within a time budget, and the loop bounds that a timing analysis needs are, in practice, often written by hand \cite{wilhelm2008worst}. Compilers already check properties of real-time code, such as the absence of blocking calls \cite{clangeffects}, but not bounds on loops and recursion. This paper describes a plugin for the Clang compiler that does: a function marked as real-time is rejected if it allocates memory, throws, calls through a pointer, or contains a loop or a recursion for which no bound is derived, and otherwise receives a bound on its cost, in abstract units with configurable weights, which can be compared with a deadline at compile time.

The analysis follows the theory of the companion paper \cite{crosscheckTheory}: loops are bounded by ranking potentials under invariants and calling contexts, recursion by depth and size potentials, and costs are composed by the timing schema \cite{shaw1989reasoning,puschner1989calculating}. Its results are of two kinds. A bound may rest on the analyzer, which is not verified; or, for functions in a linear fragment, the plugin emits a \emph{certificate}, a program annotated with invariants, potentials and Farkas multipliers, which a checker proved sound in Lean~4 \cite{demoura2021lean4,crosscheckTheory} validates without search. Only the checker and the translation from C++ to the certificate's program are then trusted, in the manner of proof-carrying code \cite{necula1997proof}. For functions in a smaller fragment, the plugin also describes the function in a small subset of C whose meaning is defined in Lean, and the checker compares that description with the certificate's program \cite[Theorem~41]{crosscheckTheory}, in the manner of translation validation \cite{pnueli1998translation}; the generator of certificates is then not trusted either.

The contributions of this paper are the design of the analysis and of the certificate generator (Section~\ref{sec:implementation}), and an evaluation on established benchmarks (Section~\ref{subsec:tacle}): loop bounds compared with flow facts and with loop counts observed by running the programs, certificates checked by the verified checker, and comparisons with five existing analyzers. The evaluation shows where the approach is weak as well as where it is useful: on programs written for complexity analyzers it bounds far fewer programs than those tools.

\section{Background}
\label{sec:background}

We recall the results of the companion paper \cite{crosscheckTheory} that the implementation relies on; proofs and Lean statements are given there.

\paragraph{Cost.} A cost norm counts the instructions of an execution trace with weights \cite[Definition~3]{crosscheckTheory}. The cost of a conditional is bounded by the cost of its test plus the maximum of its branches, and the cost of a loop with at most $K$ iterations by $K$ times the cost of its test and body plus one test \cite[Section~5.2.1]{crosscheckTheory}.

\paragraph{Ranking potentials.} Let $\to$ be a transition relation, possibly nondeterministic or partial, and $V$ a rational potential. If every step from a state of an inductive invariant $I$ decreases $V$ by at least $1$, and $V \ge 1$ whenever a step is taken, then every run from $I \cap C$ has at most $\lfloor K \rfloor$ steps, where $K$ bounds $V$ on the set $C$ of initial states (a \emph{context}); the invariant holds wherever the run stops \cite[Proposition~25]{crosscheckTheory}. The conditions on $V$ are linear in its coefficients through Farkas' Lemma, which reduces the synthesis of $V$ to a linear arithmetic problem \cite[Section~5.2.4]{crosscheckTheory}. For recursion, a depth potential bounds the depth of the call tree and a size potential $U$ with $U(x) \ge 1 + \sum_i U(x_i)$ bounds the number of activations \cite[Proposition~28]{crosscheckTheory}.

\paragraph{Cost logic and certificates.} These rules are those of a cost logic for a small language with nondeterministic atomic actions, procedures and assumed contracts, sound for terminating and non-terminating executions \cite[Theorem~31]{crosscheckTheory}. In its linear instance, a certificate is a program over integer variables annotated with intermediate assertions, loop invariants, potentials $e/d$, entry bounds and Farkas multipliers \cite[Definition~35]{crosscheckTheory}; a checker that compares integer coefficients accepts a certificate only if every execution of its program, from its precondition, costs at most the stated bound \cite[Theorem~36]{crosscheckTheory}.

\section{Prototype Implementation}
\label{sec:implementation}

The relaxed analysis of \cite[Section~5.2]{crosscheckTheory} is implemented as a plugin for the Clang compiler (about 3{,}400 lines of C++ and 1{,}100 lines of tests; directory \lean{clang-plugin} of the repository). It checks functions marked with a source annotation, and a violation is a compilation error. This section describes what the prototype does and reports a first measurement on an established benchmark collection, with a comparison to existing analyzers (Subsection~\ref{subsec:tacle}).

\paragraph{Input.}
A function is annotated as real-time, optionally with a deadline in cost units. Integer parameters may carry range preconditions, and a function may state a linear precondition relating its parameters; preconditions are assumed in the function and proved at each analysed call site. Three further annotations introduce \emph{trusted} information: a bound for a loop, a fact that holds at a program point, and the cost of a function whose body is not available. Everything trusted is listed in the report produced by the tool.

\paragraph{Profile.}
In an annotated function and in everything it calls, the plugin rejects dynamic allocation, exceptions, virtual and indirect calls, unstructured jumps, and any loop or recursion for which no bound is derived or stated. Calls are followed into their definitions, including library code.

\paragraph{Cost.}
The cost norm is a weighted count of source-level operations (arithmetic, assignments, memory accesses, calls, branches) with configurable weights; it is the weighted cost of \cite[Definition~3]{crosscheckTheory}. Costs are composed with the rules of \cite[Section~5.2.1]{crosscheckTheory}, refined so that a path ending in \texttt{return} is not charged for the code that follows it. A loop with at most $K$ iterations costs at most $K$ times the cost of guard and body, plus one failing guard (\lean{while\_\allowbreak cost\_\allowbreak bound}).

\paragraph{Bounds.}
Local integer variables whose address is not taken are mapped to affine expressions over symbols, and path conditions to linear constraints. Three kinds of values that legacy C uses for sizes are also tracked, each under an assumption that the report lists: global and \texttt{static} integer variables whose address is never taken (unknown at entry, unknown again after a call to a function that may write them; assumed not modified by interrupts or by functions without an analyzed body), fields read through a pointer parameter that the function neither writes nor changes (one symbol, their value at entry; assumed not modified through an aliasing pointer while the function runs), and entries of \texttt{const} tables with constant initializers (a value between the least and the greatest entry). Shifts, quotients, remainders and masks by constants give linear facts when the operand is provably non-negative, and integer tests in C (\texttt{while (--j)}) are disequalities. At a loop, the modified variables are replaced by fresh symbols; simple candidate invariants (a variable that is only incremented stays above its initial value; a disequality guard $a \ne b$ with $a \le b$ initially) are kept if the solver proves them inductive. The paths through guard and body, up to a fixed number, give one transition each, and Z3 \cite{demoura2008z3} is asked for an affine potential satisfying the conditions of the relational variant of \cite[Theorem~24]{crosscheckTheory} on every path, encoded through Farkas' Lemma, together with the least constant $K$ dominating the potential on entry; the loop bound is $\lfloor K \rfloor$. A logarithmic potential is also sought: a variable $v \ge 1$ under the test that every path divides by a constant $b \ge 2$ gives the bound $\lfloor \log_b K \rfloor + 1$, where $K$ bounds $v$ on entry, and so does a variable $v \ge 1$ on entry that every path multiplies by at least $b$, where $K$ bounds $v$ under the test \cite[Proposition~26]{crosscheckTheory}; the smaller of the linear and the logarithmic bound is kept. When no potential is found, the search is repeated once with more candidate invariants, each proved on entry and inductive like the others: the tests met on the paths of an iteration, as they are and relaxed by one, the range spanned by the entry value of a variable and the constants assigned to it, and equalities between variables that a path establishes. With bounds on an inner counter among the invariants, a lexicographic argument whose inner bound is a constant is an affine potential \cite[Proposition~26]{crosscheckTheory}, which the same search finds. If an iteration may start without ever returning to the loop head (an inner loop without a bound), the number of executions of the body is the number of completed iterations plus one, unless the potential is proved positive at the start of every iteration. Quotients, right shifts, remainders and masks of unsigned values are exact, since they cannot wrap. Other unsigned arithmetic is used only when it is proved not to wrap, and the limits of machine integers are never used as bounds. For recursion, both certificates of \cite[Proposition~28]{crosscheckTheory} are attempted, with the self-calls of each path as the summands, and the smaller bound is kept. After a loop, the analysis keeps one state for the exit through the guard, with the proved invariants and the negated guard, and one state for each path that leaves by \texttt{break}, with the facts of that path; this is justified by the first consequence of \cite[Proposition~25]{crosscheckTheory}.

\paragraph{Calls.}
A callee is analyzed under a calling context, following \cite[Proposition~25]{crosscheckTheory}. When a call is evaluated, the solver is asked for the interval of each integer argument in the current symbolic state; the intervals are joined over all the states in which the call is evaluated, which include every path and the generic iteration of each enclosing loop. The callee is then analyzed with these intervals as additional facts on its parameters, and the result is the cost of that call. A context is thus proved at the call site it is used for, and nothing is added to the trusted information. Functions that call themselves are analyzed without context, because the context holds for the first activation only, and the number of contexts per function is limited. The report gives, for each function, the bound that holds for every call satisfying its preconditions, and the bound that holds in the contexts through which a designated entry function reaches it. Contexts are intervals: relations between arguments, the values of global variables and the effects of a callee on memory are not propagated.

\paragraph{Examples.}
Table~\ref{tab:plugin} reports the output of the prototype on small programs written in the style of embedded firmware (directory \lean{benchmarks}). In addition, three programs are rejected as intended: one that reaches the allocator through a library container, one with a busy-wait loop, and one whose bound exceeds its deadline. The figures illustrate the behaviour of the analysis, including its pessimism: the bound for the merge sort is quadratic because every activation is charged the worst-case cost of a merge, and the recursive table lookup is given a linear depth where the true depth is logarithmic.

\begin{table}[ht]
\centering
\footnotesize
\begin{tabular}{@{}>{\raggedright\arraybackslash}p{0.21\textwidth}>{\raggedright\arraybackslash}p{0.19\textwidth}r>{\raggedright\arraybackslash}p{0.40\textwidth}@{}}
\toprule
Program & Function & Bound & Origin of the iteration or call bound \\
\midrule
FIR filter, 16 taps & one sample & 331 & 16 iterations, derived \\
 & block of $\le 64$ samples & 21\,699 & 64 iterations, from the parameter range \\
PID controller & one step & 19 & no loop; maximum over clamping branches \\
 & four axes & 167 & 4 iterations, derived \\
Range maximum (recursive) & window of 256 & 7\,154 & 511 calls, size potential $2(hi - lo) - 1$ \\
Table lookup (recursive) & window of 256 & 6\,144 & linear recursion, linear depth bound \\
Merge sort (recursive) & window of 256 & 3\,148\,271 & 511 calls, each charged a full merge \\
CAN frame decoder & checksum & 67 & 7 iterations, derived from a clamp \\
UART receive handler & drain FIFO & 309 & 16 iterations, \emph{assumed} (hardware) \\
 & read ring buffer & 838 & 64 iterations, derived \\
CRC-32, bitwise & packet of $\le 128$ bytes & 10\,502 & $128 \times 8$ iterations, derived \\
Stepper ramp planning & speed $\le 4000$ & 883 & 79 iterations, derived from the lower bound on the acceleration \\
\bottomrule
\end{tabular}
\caption{Cost bounds reported by the prototype, in abstract cost units with unit weights.}
\label{tab:plugin}
\end{table}

\paragraph{Certificates.}
For a function in the fragment of \cite[Section~5.4]{crosscheckTheory}, the plugin also emits a linear cost certificate (\cite[Definition~35]{crosscheckTheory}), which the checker, compiled from its Lean definition, validates. The certificate's variables are the tracked integer variables, signed or unsigned, that can influence control: those of tests and loop potentials, closed under the assignments and parameter passing that feed them. Every other statement is an unknown value given to a variable that no assertion mentions, at the cost of the statement. Assertions are computed forwards, with Fourier--Motzkin elimination after assignments and, after a conditional, the constraints implied by both branches; every step carries Farkas multipliers, found by Z3 over the rationals and scaled to integers. A loop uses the potential the analysis found, an invariant obtained by keeping the inductive constraints among those known on entry, and an entry bound computed by Z3. The certificates of a translation unit form one program certificate (\cite[Definition~37]{crosscheckTheory}). The bound of each function is a summary, affine in its parameters: its precondition states $x_p = \gamma_p$ for each parameter, and a loop's entry bound is expressed in the $\gamma_p$ when a Fourier--Motzkin projection of the entry assertion onto the potential and the $\gamma_p$ gives one and it is no looser, at the largest admissible arguments, than the constant one; for this, the potential suggested by the loop test ($n - i$ for $i < n$) is tried before the analysis' potential, which may use the bound of the range instead of the parameter. Inside another loop, the entry bound of a loop is first sought in the potentials of the enclosing loops (the logical variable $\gamma_j$ of each, the potential before its body): for an inner loop from the outer counter to a constant limit it is the outer potential itself. The body of the enclosing loop then has a bound $a + c\,\gamma_j$, and its loop node carries the amortised bound of \cite[Proposition~32]{crosscheckTheory}; the generator lowers the body a second time with constant inner bounds and keeps the smaller of the two certified bounds. A body that contains a loop is also lowered with $\gamma_j$ equal to $k - e$, what the potential $e$ has lost since the entry bound $k$: for an inner loop from $0$ to the outer counter, the entry bound is then $\gamma_j$ plus a constant, and the enclosing loop is certified by the loop node for costs that grow as the potential falls (\cite[Proposition~32(2)]{crosscheckTheory}), if its bound is smaller. A call of a function certified earlier is checked against its summary: the arguments are bounded by symbolic or constant bounds of the caller, and only the facts that do not mention the callee's variables are kept. A call of a function with a trusted cost, given by \texttt{CROSSCHECK\_COST} or by a contract file that states costs linear in the arguments of library functions, refers to an assumed contract, which the checker reports as such. Other calls are inlined. Several constructs are handled by transforming the program before the certificate is built, so that the checker needs no further rule: a call that is inlined is replaced by the body of the callee, analyzed in the calling context, with its parameters assigned from the arguments, which is sound because a callee cannot modify the tracked variables of its caller; the statements that follow a conditional or block that may leave the current list (by \texttt{return}, \texttt{break} or \texttt{continue}) are moved into its branches; inside loops, \texttt{break}, \texttt{continue} and \texttt{return} set flag variables, the rest of the body and the increment run only while the flags are $0$, and the loop test becomes the conjunction of the C test and ``flag $\le 0$''. Because the iteration that leaves through a flag need not decrease the potential, the certificate's potential is $e + c\,(1 - \sum \text{flags})$, where $e$ is the analysis' potential and $c$ the least multiple of the decrease, up to $64$, that the checker's conditions accept. A \texttt{switch} becomes a sequence of conditionals on an unknown value, one per group of cases (fall-through is the sequence), so that each group may or may not run; a test conjunct outside the fragment becomes an unknown value. An assignment to an unsigned variable of width $w$ becomes \emph{if $0 \le a \le 2^w - 1$ then $x := a$ else $x := {?}$}, where $a$ is an affine expression congruent to the stored value modulo $2^w$; the checker proves the wrapping branch dead wherever the assertions exclude wrap-around, so that no-wrap is not taken on trust. The weights are those of the cost model, so the certified bound is the plugin's bound, except where the forward assertions prove a path infeasible or a nest is amortised (a smaller bound) or the entry bound of a loop is rounded up (a larger one). A conditional expression whose two arms call functions is lowered with both calls in sequence, a sum where the plugin takes the maximum (a larger bound). Outside the fragment, the report states the reason: \texttt{do} loops with a jump or a \texttt{switch} in their body, \texttt{goto}, case labels nested in other statements, recursion, calls of functions with a trusted cost, and trusted annotations.

\paragraph{Validation against the source.}
With the option \texttt{minic}, the plugin also exports each certified function as a mini-C statement (\cite[Definition~39]{crosscheckTheory}). The exporter is a separate, syntactic translation of a few hundred lines that uses neither the certificate generator nor any result of the analysis, only the numbering of the certificate's variables and the cost model. A read of a tracked signed integer variable is a variable; an integer constant expression is a constant; signed sums, differences, products and negations, and conversions that preserve values, are exact, since signed overflow is undefined behaviour and is assumed absent; every other value read (memory, an untracked variable, unsigned arithmetic, a narrowing conversion) is unknown. An assignment to a tracked variable is an assignment, exact for \texttt{+=}, \texttt{-=}, \texttt{++} and \texttt{--} only when converting the result back to the variable's type preserves values; every other statement only costs. Conditionals and loops keep their structure, with the C tests ($<$, $\le$, $>$, $\ge$, negation, conjunction in loop tests). A call is a statement that stores the arguments into the variables of the callee's integer parameters and then runs the callee of the certificate file; the value of the call is unknown. \texttt{break}, \texttt{continue}, \texttt{return} and \texttt{switch} are exported as they are written, in the fragment with jumps of \cite[Definition~42]{crosscheckTheory}, whose meaning is given in Lean with outcomes; the checker itself eliminates them with flag variables, the form in which the generator writes them, and this elimination is proved (\cite[Theorem~43]{crosscheckTheory}). The exporter only names the flag variables of the certificate (and, for a \texttt{switch}, the variable of the unknown choice of the case group), which it finds by name; the checker refuses names that are variables of the source or are not fresh, and it checks that no function that the validated function can call touches them. A \texttt{switch} enters one of its case groups, which one is not modelled, and falls through to a \texttt{break}: every execution of the C statement is one of its executions. Tracked global variables are variables of the certificate shared by the functions, read and written as locals are; the hypothesis is that no other code (an interrupt handler, another thread) writes them while the function runs. The exporter refuses the function, and says why, for calls in tests, calls of functions that have no entry of their own in the file, \texttt{goto}, tracked variables that are static locals, volatile, of an unsigned type narrower than \texttt{unsigned int}, or whose address is taken, and assignments of a constant-table read; the checker refuses a loop without a test of its own that is left by a jump. A \texttt{do} loop is exported as such; its meaning is also in Lean, and the checker accepts it when its body has no \texttt{break} or \texttt{continue} of its own, as its body followed by the loop; the generator certifies a \texttt{do} loop in that form when its body has no jump and no \texttt{switch} at all. A quotient, remainder, right shift or mask by a constant is a statement of its own, whose value is described by the facts of the definition; the exporter checks on the C types that the constant is non-negative and that the result is stored without conversion. A store into an unsigned variable is the statement that keeps the value only if it is in the variable's range; its right-hand side is taken over the integers when no intermediate result is truncated to fewer bits than the variable has. The checker lowers the statement with the Lean definition of mini-C and compares it with the certificate's program, without annotations; for this, the generator puts the code of an infeasible branch inside the dead-code node instead of an empty statement. A certificate that is accepted and validated bounds every execution of the mini-C description of the function.

\paragraph{Trust.}
The implementation is not verified. What is machine-checked is the mathematics behind each kind of certificate it produces (\cite[Section~7]{crosscheckTheory}). For a function with an accepted certificate, the bound itself no longer depends on the analysis, its invariants or the SMT solver: it depends on the checker, which is proved sound, on the assumed contracts of the functions it calls with a trusted cost, which are listed, and on the translation from the C++ function to the certificate's program, which is not verified. For a validated certificate, that translation is replaced by the export to mini-C, which is shorter, syntactic and independent of the analysis, together with the hypotheses under which a C function means its description (no signed overflow, no aliasing of tracked variables, no concurrent writes). For the other functions the translation from the Clang syntax tree to constraints, the invariant generation and the SMT solver are trusted. The cost unit is not a duration (Section~\ref{sec:limitations}).

\paragraph{Rust.}
A front-end for Rust translates the MIR of a crate, as printed by \texttt{rustc} with optimizations, overflow checks and \texttt{panic=abort}, into C, which the plugin then analyzes and certifies as any C code. MIR is a control-flow graph; loops are rebuilt as \texttt{while} loops whose test is the conjunction of the tests at their head, and conditionals use immediate post-dominators. The fragment is that of free functions over integers, slices of integers and calls between them; \texttt{for} loops over ranges are covered because the optimizer inlines their iterator into integer code, while slice iterators, loops left by \texttt{break} and structured types are not. Wrapping operations are computed in C on unsigned types. Checked arithmetic, which panics on overflow, becomes plain C arithmetic, which the analysis assumes not to overflow: the bound covers the executions that do not panic. Every other assertion (bounds checks, user \texttt{assert!}) and every call that does not return becomes an early return, which gives the C program more behaviours than the Rust one and makes the asserted fact available afterwards; an \texttt{assert!} on an input is thus a range that is checked at run time instead of trusted. The textual MIR is meant for humans and changes between Rust releases: the translator is written for one version of \texttt{rustc}, and a stable interface would require a compiler driver on a nightly toolchain. The translation is checked by a differential test: on a set of 12 small kernels, the 11 translated functions agree with Rust on 382 random inputs (Rust panics skipped), and the 4 functions that must not be bounded (Euclid's algorithm, the Collatz iteration, a wrapping loop that never exits, a loop that ends only by overflow) are not; 6 functions are bounded, with \texttt{assert!} ranges and no annotation, and 3 of them certified. The translation itself is not verified.

\paragraph{Interrupts.}
A function annotated \texttt{CROSSCHECK\_INTERRUPT(T)} is an interrupt handler activated at most once every $T$ cost units, a trusted value listed among the assumptions. Handlers are analysed like real-time functions, and every other analysed function receives, in addition to its cost bound $C$ (in context, when the program's calling contexts give one), a response-time bound: the least $R$ with $C + \sum_k (c_k + o)\lceil R/T_k\rceil \le R$, where $c_k$ is the bound of handler $k$ and $o$ a per-activation overhead given by an option (\cite[Theorem~46]{crosscheckTheory}). A deadline is then checked against the response time. When the function and the handlers have certificates with constant bounds, the certificate file also contains the claim with the certified bounds, which the checker verifies. The assumptions are those of the theorem: no activation pending when the function starts, minimum inter-arrival times, the overhead, and that a cost unit is a time unit, which holds only for a cost model of the target (Section~\ref{sec:limitations}). Handlers that write variables the function's bounds depend on are not detected: global variables are assumed not to be written by interrupts. Fork--join parallelism (\cite[Proposition~47]{crosscheckTheory}) is not supported by the plugin.

\paragraph{Exports.}
The derived loop bounds can be written as flow facts for WCET analyzers, in two formats. In FFX, the XML format of OTAWA, written by oRange \cite{bonenfant2012ffx}, a loop is designated by its source file and line and receives a \texttt{maxcount}, the maximal number of iterations per entry of the loop. In the AIS and AIS2 annotation languages of aiT \cite{heckmann2013ait}, a loop is designated by the linker name of its function and a loop number. aiT numbers the loops of a routine in the executable, while the plugin numbers them in source order; the two agree only if the compiler neither reorders, unrolls nor inlines the loops, which the file states and the user must check. The exported bound is either the one that holds for every call or the one that holds in the calling contexts of the program. Bounds stated by annotations are assumptions and are not exported; the analysis' other assumptions are listed in the file's header. The results can also be written in SARIF \cite{sarif2023}, the interchange format of code-scanning interfaces: rejected functions and missed deadlines as errors, cost bounds and assumptions as notes. For continuous integration, a script compares the bounds of two versions of a program function by function and fails when a function is newly rejected or a bound grows beyond a budget. The FFX and SARIF outputs are read back in the tests, the latter validated against the SARIF schema; neither format has been fed to OTAWA or aiT, which need compiled binaries. The reports also give, for every loop, the invariants proved at its head and a variant (the potential scaled to integers, or the divided or multiplied counter of a logarithmic loop), and for every function its preconditions on integer parameters; a script writes them into a copy of the source as ACSL annotations, the specification language of Frama-C \cite{kirchner2015framac} (\texttt{requires}, \texttt{loop invariant}, \texttt{loop variant}), for a deductive verifier to prove independently. The annotations are tested by executing them (paragraph ``Tests against executions'' below); they have not been given to a deductive verifier. A driver runs the tool on every translation unit of a compilation database and produces one report for the project. An unbounded loop is reported with a likely cause: a variable of its test without a range where the loop starts, a test that reads memory or a floating-point value, or bounded variables without a decreasing combination.

\paragraph{Time budgets.}
The cost bounds are in abstract units, while industrial budgets are durations, a property of the machine code on a processor: an annotation \texttt{CROSSCHECK\_DEADLINE\_US(n)} (or \texttt{\_CYCLES}) is only recorded in the report, together with the loop bounds, and a script checks it on the binary after compilation, with a timing model of the processor \cite{crosscheckBinary}.

\subsection{TACLeBench kernels}
\label{subsec:tacle}

TACLeBench \cite{falk2016taclebench} is a collection of self-contained C programs for WCET research, gathered from several earlier suites, among them the M\"alardalen WCET benchmarks \cite{gustafsson2010malardalen}. Every loop of its programs carries a flow fact \texttt{\#pragma loopbound min $X$ max $Y$}. We ran the tool on the 29 programs of its \emph{kernel} category (revision \texttt{4e4fbf7} of the public repository), analysing every function, \emph{without adding any annotation}; the files of each program are concatenated into one translation unit. For each loop with a flow fact, Table~\ref{tab:tacle} compares the bound derived by the tool, if any, with the maximum $Y$ of the flow fact. Two bounds are reported for a loop: the one that holds for every call of its function, and the one that holds in the calling contexts through which the program's \texttt{main} function reaches it (paragraph ``Calls'' above); the second is the one comparable with the flow facts, which describe the program on its fixed input. The script that fetches the benchmarks and produces the table is \lean{benchmarks/tacle/run\_tacle.py}; the whole run takes about four minutes on a laptop (about two and a half before the search for a potential was repeated with more candidate invariants).

\begin{table}[ht]
\centering
\footnotesize
\setlength{\tabcolsep}{4pt}
\begin{tabular}{@{}lrrrrrrrrr@{}}
\toprule
 & & \multicolumn{6}{c}{For every call} & \multicolumn{2}{c}{In context} \\
\cmidrule(lr){3-8}\cmidrule(l){9-10}
Kernel & Loops & Derived & $= Y$ & $> Y$ & $< Y$ & None & \texttt{\_main} & Derived & $= Y$ \\
\midrule
binarysearch & 2 & 2 & 1 & 1 & 0 & 0 & 386 & 2 & 1 \\
bitcount & 4 & 2 & 1 & 1 & 0 & 2 & -- & 2 & 1 \\
bitonic & 3 & 2 & 2 & 0 & 0 & 1 & -- & 2 & 2 \\
bsort & 4 & 4 & 4 & 0 & 0 & 0 & 240\,702 & 4 & 4 \\
complex\_updates & 4 & 4 & 4 & 0 & 0 & 0 & 871 & 4 & 4 \\
cosf & 3 & 1 & 1 & 0 & 0 & 2 & -- & 0 & 0 \\
countnegative & 4 & 4 & 4 & 0 & 0 & 0 & 5\,752 & 4 & 4 \\
cubic & 6 & 1 & 1 & 0 & 0 & 5 & -- & 1 & 1 \\
deg2rad & 1 & 0 & 0 & 0 & 0 & 1 & -- & 0 & 0 \\
fac & 1 & 0 & 0 & 0 & 0 & 1 & -- & 0 & 0 \\
fft & 12 & 11 & 8 & 3 & 0 & 1 & 271\,351\,277 & 11 & 8 \\
filterbank & 14 & 11 & 11 & 0 & 0 & 3 & -- & 11 & 11 \\
fir2dim & 17 & 9 & 9 & 0 & 0 & 8 & -- & 9 & 9 \\
iir & 6 & 6 & 6 & 0 & 0 & 0 & 210 & 6 & 6 \\
insertsort & 4 & 2 & 2 & 0 & 0 & 2 & -- & 2 & 2 \\
isqrt & 5 & 3 & 3 & 0 & 0 & 2 & 861\,860 & 3 & 3 \\
jfdctint & 4 & 4 & 4 & 0 & 0 & 0 & 2\,699 & 4 & 4 \\
lms & 7 & 4 & 4 & 0 & 0 & 3 & 115\,950 & 7 & 7 \\
ludcmp & 12 & 12 & 3 & 9 & 0 & 0 & 4\,903 & 12 & 12 \\
matrix1 & 7 & 7 & 7 & 0 & 0 & 0 & 15\,496 & 7 & 7 \\
md5 & 9 & 2 & 2 & 0 & 0 & 7 & -- & 5 & 5 \\
minver & 21 & 17 & 8 & 9 & 0 & 4 & -- & 20 & 20 \\
pm & 30 & 5 & 5 & 0 & 0 & 25 & -- & 5 & 5 \\
prime & 1 & 0 & 0 & 0 & 0 & 1 & -- & 0 & 0 \\
quicksort & 16 & 5 & 4 & 0 & 1 & 11 & -- & 5 & 4 \\
rad2deg & 1 & 0 & 0 & 0 & 0 & 1 & -- & 0 & 0 \\
recursion & 0 & 0 & 0 & 0 & 0 & 0 & -- & 0 & 0 \\
sha & 17 & 10 & 6 & 4 & 0 & 7 & -- & 11 & 6 \\
st & 5 & 5 & 5 & 0 & 0 & 0 & 65\,074 & 5 & 5 \\
\midrule
Total & 220 & 133 & 105 & 27 & 1 & 87 & 12 of 29 & 142 & 131 \\
\bottomrule
\end{tabular}
\caption{The prototype on the TACLeBench kernels, without annotations. Loops: loops with a flow fact. For every call: loops for which a bound is proved that holds for every call of their function, split by comparison with the maximum $Y$ of the flow fact, and cost bound of the entry function \texttt{<kernel>\_main}, in abstract units with unit weights (-- when a loop or a call it reaches is not bounded). In context: loops bounded in the calling contexts through which the program's \texttt{main} reaches them.}
\label{tab:tacle}
\end{table}

A bound that holds for every call is derived for 133 of the 220 loops, and 105 of these bounds equal the maximum of the flow fact. The 27 larger bounds are mostly bounds of loops controlled by a parameter, which hold for every admissible value of the parameter (in \texttt{ludcmp} and \texttt{minver}, 99 and 500 for sizes that the program sets to 5 and 3); the others include three loops of \texttt{fft}: a halving and a doubling loop, which receive the logarithmic bound 11 where the flow fact states 10, and a loop of step 2, bounded by 1023 where it states 512, and the loop of \texttt{binarysearch}, which receives a linear bound, 29, where the flow fact states 4: its potential is a difference of two indices, found since the bounds of the indices are among the candidate invariants, but it is not recognised as logarithmic. In context, 142 loops are bounded and 131 bounds equal the flow fact: the loops of \texttt{ludcmp}, \texttt{minver} and \texttt{lms} receive the sizes passed by their callers, and the cost bound of \texttt{ludcmp\_main} drops from 22\,961\,019 units to 4\,903. The one smaller bound is a loose flow fact: the inner loop of \texttt{quicksort\_init} is \texttt{for (j = 0; j < 20 - 1; j++)}, which has at most 19 iterations, while the flow fact states 20. Flow facts describe each program on its fixed input; they are a reference, not a proof of the worst case. The 87 loops without a bound for every call (78 in context) are classified by the cause that the analyser states in its diagnostic (\lean{benchmarks/tacle/classify\_unbounded.py}): for 55, a variable of the test has no known range where the loop starts (a parameter whose range is not known at the call either, a global variable, a structure field, or a value read from memory; the guard $i \cdot i \le n$ of \texttt{prime} is among them); for 27, the test does not depend on a tracked integer variable: 17 have a floating-point or \texttt{volatile} counter, which the analyser does not model, and 10 are driven by the data (partitioning, string comparison, pointer comparison, \texttt{while (1)} loops left by \texttt{break}); for 5, the variables of the test are bounded but no potential is found (a limit that is a global variable or a floating-point value, a test combined with \texttt{\&}, insertion sort). The split of the 27 is by reading the sources. The two recursive kernels are rejected because the recursion argument has no range. For 12 of the 29 kernels the entry function receives a cost bound with no annotation and no assumption. Of the 165 functions that receive a bound, 161 also receive a certificate that the checker accepts, including the entry functions of 11 kernels; 146 certified bounds equal the plugin's and 12 are smaller: in 9 an infeasible path is excluded (\texttt{\_\_kernel\_cosf} and its callers, in four kernels), in 3 (\texttt{bsort}) the iteration that leaves by \texttt{break} is counted without the increment and without the extra body of the plugin's rule. 3 are larger: the entry bound of a loop is rounded up (\texttt{ludcmp}, 2), and a conditional expression calls a function in each arm (\texttt{pm\_init\_data}). The four other functions receive no certificate: one contains an assignment inside an expression, one a loop whose potential has no constant upper bound on entry in the certificate's assertions, and in two functions of \texttt{fft}, bounded since logarithmic potentials were added, the generator's invariant does not show the logarithmic potential positive. Written as FFX, the in-context bounds of the kernels agree with the FFX files produced by oRange on 115 of the 116 loops that both bound numerically, the exception being the loop of \texttt{binarysearch} (29 against oRange's 5) (all loops of the files, not only those with a flow fact; \lean{benchmarks/tacle/compare\_ffx.py}). Before early exits, \texttt{switch}, conjunctive tests and unsigned variables were supported by the generator, 142 functions were certified; these constructs account for the difference. 128 calls are checked against the callee's certificate instead of being inlined; no kernel function has a summary that depends on its parameters, as the kernels' loops have constant bounds. With the option \texttt{minic}, 152 of the 161 certificates are validated against the mini-C description of their function (93 before quotients, remainders, shifts and masks by constants, stores into unsigned variables, calls, early exits, \texttt{switch} and tracked globals were covered); 146 are validated together with every function they call. The certified bound equals the plugin's for 146 certificates and is lower for 14; among these, the amortised loop rule of \cite[Proposition~32]{crosscheckTheory} lowers the bound of \texttt{ludcmp\_main} from 4{,}907 to 3{,}474 units: its three triangular nests are amortised, one with an inner loop from the outer counter to the limit and two with an inner loop whose length grows as the outer potential falls (3{,}674 with the first form only). In the sequential and application programs it lowers that of \texttt{susan\_median}, a sort of eight values in a triangular nest, from 1{,}047 to 648 (validated); no other certificate uses the rule. An outer body whose cost is quadratic in the outer counter because it contains both shapes (the first nest of \texttt{ludcmp\_test}, amortised in one dimension only) and entry bounds in a parameter are not covered; why the nest of \texttt{gsm\_enc} whose inner loop runs to $8 - n$ is not amortised was not analysed. 37 of the validated functions contain loops (61 of the 83 loops of the certified programs), and 8 are entry functions of kernels, 7 of them with all their callees (\texttt{binarysearch}, \texttt{bsort}, \texttt{complex\_updates}, \texttt{countnegative}, \texttt{iir}, \texttt{jfdctint}, \texttt{matrix1}). The exporter refuses 8: calls of functions without an entry of their own in the certificate file (6) and calls in tests (2). One certificate, of \texttt{isqrt\_main}, is exported and not validated: the generator inlined a call whose argument (a global) has no bound, so that the callee's summary could not be used, while the source describes the call as a call; the checker refuses it, as it should. Many of the validated functions are small (initialisations, the functions that return a kernel's result).

\paragraph{With minimal annotations.}
The kernels above are analyzed without any annotation. To measure what an engineer would have to add, the script \lean{benchmarks/tacle/run\_annotated.py} takes the loops that the analyzer reports as unbounded in the functions reached from \texttt{main} (the C entry point, which calls the kernel's initialization, its entry function and its result check), gives each the trusted annotation \texttt{CROSSCHECK\_LOOP\_BOUND(}$n$\texttt{)} with $n$ the maximum of the benchmark's flow fact for that loop, and repeats until \texttt{main} is bounded or no such loop is left. Without annotations \texttt{main} is bounded in 10 of the 29 kernels; with 84 annotations in 18 kernels, it is bounded in 23. In the 6 others what is missing is not a loop bound: a recursive function in five (\texttt{bitcount}, \texttt{bitonic}, \texttt{fac}, \texttt{quicksort}, \texttt{recursion}), and in \texttt{lms} two loops without a flow fact. These annotations are assumptions: the report lists them, and a function whose bound rests on one receives no certificate. Each annotated value was compared with the loop counts observed on the built-in input: two are one below, in \texttt{md5} and \texttt{quicksort}, the flow facts of two \texttt{while (1)} loops left by \texttt{break}, which count the iterations that return to the loop head (see the comparison with observed counts below), so the bound of \texttt{md5}'s \texttt{main} rests on a wrong assumption, which the comparison detects.

\paragraph{Sequential and application programs.}
The 23 programs of TACLeBench's \emph{sequential} category and the 2 of its \emph{app} category are larger (up to about 10{,}000 lines each) and were run in the same way. Of their 572 loops with a flow fact, 275 are bounded for every call and 258 of these bounds equal the flow fact; in context, 330 are bounded and 304 equal. Nine bounds for every call and ten in context are below the flow fact, none of them a counterexample: two loops of \texttt{h264\_dec} run exactly \texttt{sizeof} of an array (8{,}450 and 8{,}100 iterations, the derived bounds), where the flow facts state 33{,}800 and 16{,}200, and a loop of \texttt{lift} runs 13 times where its flow fact states 14; seven loops lie in branches that cannot be taken (bound $0$): six in \texttt{h264\_dec}, in the \texttt{else} branch of a test that an assignment just before makes true or under \texttt{if (smb)} with \texttt{smb} set to 0 and never changed, and one in \texttt{cjpeg\_wrbmp} on a branch that no call of the program reaches. Loop counts were not observed on these programs. The search for invariants is costly on large functions with nested loops that have no bound: the slowest program, \texttt{cjpeg\_transupp}, took 2{,}277 seconds of analysis before solver answers were cached and the search budget was introduced (Section~\ref{sec:limitations}), and 376 seconds after, with the same bounds; the two categories took 23 minutes in all, against 88. The 297 loops without a bound mostly depend on values read from memory, global or structure fields, pointers or floating-point values, which the analysis does not track. The entry functions of 4 of the 25 programs are bounded. Of the 263 functions that receive a bound, 231 receive a certificate, all accepted by the checker (207 with the plugin's bound, 14 smaller; 165 calls checked against the callee's certificate); the others contain assignments inside expressions (17, mostly in the GSM codecs), loop tests or conditions with a call or an assignment (10), or calls of library functions with a trusted cost (5). With the option \texttt{minic}, 194 of the 231 certificates are validated against the mini-C description of their function (113 before the extensions of mini-C, 178 before tracked globals, 193 before static locals), including the three certified entry functions (of \texttt{adpcm\_dec}, \texttt{adpcm\_enc} and \texttt{petrinet}), and none fails validation; the exporter refuses the others for divisions or shifts outside the modelled forms (12), calls of functions without an entry of their own in the certificate file (12), variables of narrow types (9), calls in tests (3) and a table read (1). One function was exported and not validated when this was first run: a constant expression, \texttt{8192 >> 1}, was exported as a quotient where the generator had folded it; the exporter now folds constants first.

\paragraph{Industrial programs.}
TACLeBench also contains three programs derived from industrial systems, in its parallel category: ROSACE, a flight-controller case study (aircraft dynamics, filters and altitude and vertical-speed control laws, in five threads, each a program), DEBIE, the software of an instrument flown on a satellite and the benchmark of the WCET Tool Challenge, and PapaBench, a UAV autopilot. PapaBench does not compile as distributed (a header is missing) and is left out; the other two need small fixes to their text (a function name that clashes with the C library, a pragma inside a function header, an include guard with a typo), applied by the script to a copy. In each ROSACE thread 73 of 86 functions are bounded and certified, among them every task of the control loop except the aircraft dynamics, which call the single-precision sine and cosine of the bundled mathematical library, whose argument reduction has loops without a bound for an arbitrary argument; the scheduler loop calls the tasks through function pointers, which the plugin does not follow. In DEBIE 111 of 187 functions are bounded and 110 certified, and 30 of the 76 loops that carry a flow fact are bounded, 26 equal to it; two bounds are below the flow fact, for loops that run once per byte of a structure that has 120 bytes on the host the code was analysed for and 124 in the build the flow facts come from. 29 of the 46 loops without a bound are in the test harness. All 480 certificates are accepted, and 309 are validated against their source (189 before static local variables were exported); one more is exported and not validated: in \texttt{Switch\_SU\_State} of DEBIE the certificate generator treats a structure field read through a pointer as a variable, while the exporter describes the read as an unknown value, so the checker refuses the match (the certificate itself is accepted). These programs show the same picture as the sequential ones at a larger scale: control code with constant loop bounds is bounded and certified, and most losses are floating-point library code, indirect calls and the test harness; the 14 loops without a bound in DEBIE's own software were not analysed one by one.

\paragraph{An embedded library.}
To measure the coverage of code written for embedded targets rather than for benchmarks, we ran the tool on CMSIS-DSP, Arm's signal-processing library for Cortex-M processors (Apache-2.0; 632 source files, 786 functions, 1{,}235 loops), for a Cortex-M4 target (\lean{benchmarks/cmsis/run\_cmsis.py}). As is, with each file alone and no annotation, 161 functions are bounded: the library's sizes are parameters or fields of instance structures, without ranges. With every integer input without a declared range assumed in $[1, 1024]$ (parameters and fields; an assumption of the run, listed in the reports), the other files of the same group as context, and illustrative contracts for \texttt{memset} and \texttt{memcpy}, 616 functions (78\%) and 1{,}069 loops are bounded (1{,}044 in two earlier runs on a shared machine, where one or two files exceeded the time limit; the number of bounded functions is the same). Before the tracking of fields, globals and C tests and the cost rules for the Arm intrinsics, statement expressions and one-instruction inline assembly of the CMSIS headers were added, the same run with ranges on parameters only bounded 357 functions. Logarithmic potentials bound 20 more loops of the transforms than linear ones did (1{,}024 before), but no more functions: the remaining blockers are mostly loops controlled by values read from memory and calls into other parts of the library outside the context.

\paragraph{Tests against executions.}
The analysis is not verified, and a certificate covers only the functions of the fragment, so the bounds were also tested against executions beyond the benchmark inputs. A generator produces random functions over \texttt{int} or \texttt{unsigned} parameters with small declared ranges, whose loops are drawn from the shapes the analysis is meant to bound (counters, steps, halving, doubling, reset of an inner counter, early exits) and from arbitrary combinations of tests and updates. Each file is analysed, then compiled with counters in its loops and run on all the inputs of its ranges, or on a sample that includes the corners; runs in which a signed value overflows are discarded, since the analysis assumes that they do not happen. In a campaign of 120 files (960 functions, 1{,}632 loops, 1.23 million runs kept), 863 loops received a bound; none was below an executed count, and 731 bounds were equal to the largest count observed. 221 certificates were emitted for these functions and all were accepted. The same runs execute the invariants and variants exported for ACSL, as checks at the start of every loop body, and compare the mini-C export of each function, run by an interpreter written from the Lean definition, with the compiled C on the same inputs: 2{,}114 invariants and 832 variants held on every run, and the 143 functions whose export could be compared counted the same loop iterations as the C code. Three exported functions were not validated by the checker in this campaign, because the exporter was less precise than the generator on a left shift of an unsigned variable and associated scratch variables with statements in another order; both were corrected, and validation, as intended, had refused them; a further campaign of 40 files (320 functions, 283 bounded loops) then had no failure of any kind. When these tests were first run they found two defects, since corrected: a loop whose body can never return to the loop head (it contains a loop that does not terminate) was reported with $0$ executions of its body, where the body starts once; and the exporter wrote a mini-C source for a loop test with a disequality, which mini-C cannot express, so that the checker refused to validate it. The first was a wrong loop bound in a function that the tool rejected in any case; neither affected a certified bound. A third defect, found by reading the code, did: the declaration of a static local variable was translated as an assignment at every call, whereas C initialises it once, so that a loop bounded by a static counter that grows from call to call received a bound, and a certificate, accepted by the checker, of a program that is not the function's. Validation against the source would have refused it, since the exporter did not translate static locals; it has been corrected (the declaration does nothing at a call, and the value on entry is unknown, as for a global), with a test that the loop is now rejected. It is an instance of the risk stated above: the translation from C to the certificate's program is trusted unless the certificate is validated. The checker program was tested separately: one integer of an accepted certificate is changed, and a mutant that the checker still accepts is run by an independent interpreter of certificate programs, on random states of its precondition. Of 2{,}029 mutants of 15 accepted programs, 1{,}983 were rejected; the 46 accepted ones remained true on every run (for instance a larger bound, or other valid multipliers). This covers the part of the checker that the proof does not, its reader of certificate files.

\paragraph{Machine code.}
The loop bounds of this section are also used for cycle and stack bounds of the kernels compiled for a Cortex-M0, certified by the checker on the machine code and compared with emulated runs and with the WCET analyzer OTAWA, and for a check of time budgets after compilation; these, and certificates of the loop bounds on the machine code itself, are described in a third paper \cite{crosscheckBinary}.

\paragraph{Comparison with other tools.}
We ran five existing analyzers on the same files, without annotations: the loop trip-count analysis of LLVM \cite{lattner2004llvm} (\emph{ScalarEvolution}, version~18), which gives a constant maximum number of iterations per loop, intraprocedurally; KoAT2 (release~2.1.0), which bounds the runtime complexity of an integer program, here translated from C by llvm2kittel \cite{falke2011termination}, by computing runtime bounds and size bounds of the transitions alternately, for one part of the program at a time \cite{brockschmidt2016analyzing}, with extensions such as bounds for triangular weakly non-linear loops \cite{lommen2022automatic}; CoFloCo, which bounds the cost of the same integer programs, converted by its own front-end into cost equations, non-deterministic recurrences with linear constraints in which loops and linearly recursive procedures are treated alike: it refines the equations of each strongly connected component into sequences of phases, infers invariants for them and combines the bounds of the phases \cite{floresmontoya2014cofloco}; the cost analysis of Infer \cite{infercost} (version~1.3.0), which computes a symbolic cost per function, interprocedurally; and oRange \cite{demichiel2008orange}, the loop-bound analyzer of the OTAWA WCET tool chain, which works on C source: it first rewrites the program into structured C without \texttt{goto}, \texttt{switch}, \texttt{break}, \texttt{continue} and (by its documentation) multiple returns, with FrontC's Calipso, then bounds the loops by flow analysis and abstract interpretation, on a context tree in which every call is expanded (no recursion), of the whole program from its entry point or of functions in isolation (we used the whole-program mode; built from its source repository, last changed in 2022). oRange reads \texttt{loopbound} pragmas, so we removed them from the files given to it. The cost units of CrossCheck, KoAT2 and Infer differ, so at the level of functions we compare only whether a finite bound is found; KoAT2 and CoFloCo count the same unit (transitions of the same integer program), so their bounds can also be compared with each other. Table~\ref{tab:tools} summarizes the results; the per-kernel figures, the answers of every tool and the scripts are in the directory \lean{benchmarks/tacle} of the repository.

\begin{table}[ht]
\centering
\footnotesize
\setlength{\tabcolsep}{3.5pt}
\begin{tabular}{@{}lrrrrr@{}}
\toprule
 & \multicolumn{2}{c}{CrossCheck} & & & \\
\cmidrule(lr){2-3}
All 220 loops & every call & in context & SCEV & oRange & fact \\
\midrule
bounded & 133 & 142 & 113 & 131 & \\
\quad $= Y$ / $> Y$ / $< Y$ & 105 / 27 / 1 & 131 / 9 / 2 & 108 / 5 / 0 & 127 / 3 / 1 & \\
by the integer type only & -- & -- & 46 & -- & \\
\midrule
The 208 loops reached & & & & & \\
\midrule
bounded & 131 & 142 & 113 & 131 & 208 \\
\quad $<$ / $=$ / $>$ observed & 0 / 103 / 28 & 0 / 132 / 10 & 0 / 107 / 6 & 1 / 127 / 3 & 2 / 198 / 8 \\
\quad mean bound / observed & 1.85 & 1.04 & 1.02 & 1.00 & 1.01 \\
\midrule
Functions & CrossCheck & KoAT2 & CoFloCo & Infer & \\
\midrule
29 entry functions, numeric & 12 & 5 & 8 & 14 & \\
entry functions, symbolic & 0 & 5 & 0 & 2 & \\
all 267: numeric / symbolic & 165 / 0 & 119 / 23 & 128 / 11 & 190 / 14 & \\
\bottomrule
\end{tabular}
\caption{The TACLeBench kernels, without annotations: loop bounds of CrossCheck, LLVM's ScalarEvolution and oRange compared with the flow facts (maximum $Y$) and with the largest number of body executions per entry of the loop observed on the kernel's input (the mean is geometric, over the loops with a nonzero count), and functions for which CrossCheck, KoAT2, CoFloCo and Infer find a finite bound. For CrossCheck, ``every call'' is the bound that holds for every call of the loop's function and ``in context'' the bound in the calling contexts reached from \texttt{main}; SCEV is ScalarEvolution and ``fact'' the flow fact. A symbolic bound depends on values the tool does not know.}
\label{tab:tools}
\end{table}

On loops, CrossCheck's bounds for every call and ScalarEvolution's, both obtained from one function at a time, are close. 104 loops are bounded by both; the bounds agree on 99, on 4 CrossCheck's is smaller by one (loops left by \texttt{break}, where the iteration count is one more than LLVM's backedge-taken count), and on one, the doubling loop of \texttt{fft}, ScalarEvolution's is smaller by one (10, the exact count, against the logarithmic bound 11). 29 loops are bounded only by CrossCheck, 18 of them in \texttt{ludcmp} and \texttt{minver}, where ScalarEvolution obtains no bound or only the range of the integer type, which CrossCheck never uses. 9 are bounded only by ScalarEvolution: loops with floating-point counters and constant start, step and limit, which it evaluates exhaustively. To check the loop bounds against executions, and not only against the flow facts, we instrumented the unoptimized code of each kernel to record, for every loop, the largest number of executions of its body in one entry of the loop, and ran it on its built-in input; 208 of the 220 loops are reached. No bound of CrossCheck, for every call or in context, and no bound of ScalarEvolution is below the observed count. The flow facts are equal to it on 198 loops, and below it by one on two loops of the form \texttt{while (1)} left by \texttt{break}, which is consistent with counting only the iterations that return to the loop head. oRange propagates the values of the whole program from its entry point, so its bounds are bounds for the program with its built-in input, like the flow facts and like CrossCheck's bounds in context: in \texttt{ludcmp} and \texttt{minver} both obtain the sizes 5 and 3 with which the kernel calls the function, where the bounds for every call are 99 and 500. The two agree on 114 of the 115 loops that both bound; on the loop of \texttt{binarysearch} oRange's bound is 5 and ours 29. 27 loops are bounded by CrossCheck in context and not by oRange: 10 in the five kernels on which oRange gave no result (it rejects three because of their recursion, raised an internal error on one, and did not finish within 600 seconds on one), and 17 in \texttt{fft}, \texttt{fir2dim}, \texttt{iir}, \texttt{isqrt}, \texttt{md5} and \texttt{sha}, six of which have a guard of the form \texttt{i < sizeof(a)}. 16 loops are bounded by oRange and not by CrossCheck: loops with floating-point counters (12, eight of them in \texttt{fir2dim}), and loops controlled by a global or \texttt{volatile} variable or by the data. One oRange bound is below the observed count: in \texttt{filterbank}, the loop \texttt{while (n-- > 0)} with $n = 2$ runs twice (the flow fact also states 2), and oRange reports at most one iteration. The instrumentation is tested on a program with known counts, this form of loop included (\lean{benchmarks/tacle/test\_measure\_loops.py}). The observed counts hold for one input only, so a bound above them is not necessarily loose; for the bounds that hold for every call, the ratios mainly measure the distance between all admissible calls and the calls the program makes (the largest, $500/3$ in \texttt{minver}, comes from the test $\mathit{side} \le 500$ at the start of a function that the kernel calls with $\mathit{side} = 3$), and in context the geometric mean drops from 1.85 to 1.04. At the level of functions, Infer finds the most bounds, mainly because it is interprocedural: a loop bounded by a parameter is bounded at the call site. Its results are not upper bounds in every case: by its documentation \cite{infercost} the analysis does not handle recursion, and it reports a constant cost for the recursive Fibonacci kernel. KoAT2 is sound for the integer program it analyzes, but the translation keeps only integer variables, so values in memory become unknown inputs (hence its symbolic bounds), and for 88 functions the translation or the analysis failed or timed out; it is the only tool to bound the kernel \texttt{prime}, whose guard $i \cdot i \le n$ is not affine. On the same 202 integer programs, CoFloCo finds more numeric bounds than KoAT2 (128 against 119) and fewer symbolic ones; of the 117 functions that both bound by a number, the bounds are equal for 97 and CoFloCo's is smaller for 20 (e.g.\ 40{,}798 against 59{,}900 transitions for \texttt{bsort\_main}), never larger. Its entry-function bounds cover the same kernels as CrossCheck's, except \texttt{st}, on which it timed out. No function bounded by CrossCheck is reported unbounded by KoAT2 or CoFloCo (where they fail, the translation failed or they timed out); Infer finds no bound for one of them, whose loop CrossCheck bounds by the guard $i < 64$. The tools answer different questions (constant bounds for real-time code, asymptotic complexity, cost regressions), and none was given annotations or tuned; the comparison shows where each fails, not which is better in general.

\paragraph{Complexity benchmarks.}
TACLeBench favours CrossCheck, since its loops have constant bounds. Complexity analyzers are evaluated instead on the C programs of the complexity category of the Termination and Complexity Competition \cite{giesl2019termcomp} (directory \lean{Complexity\_C\_Integer} of the Termination Problem Database, 553 programs at the revision we used): functions of integer parameters, often with nondeterministic values, whose costs are symbolic. The analysis derives numeric bounds only, so on the unchanged programs it bounds 11, those whose loops have constant bounds. With the option \texttt{symbolic}, a loop needs only a ranking potential, which requires no range: the analysis then gives no number, and the certificate generator derives a bound in $\max(0, x)$, $\max(0, -x)$ and $\max(0, x - y)$ for the integer parameters $x, y$ of the function, affine or quadratic (a product of two of these terms, for a nest of two loops over parameters, \cite[Section~5.4]{crosscheckTheory}), which the checker accepts or rejects. In this mode a value that the entry function reads from the nondeterministic input function before its first loop is made a parameter by the harness (the other tools treat these values as inputs too). 147 of the 553 programs then receive a certificate that the checker accepts, 133 with a bound in the parameters and 14 with a constant one: bounds that hold for all inputs, without annotations. All 147 are also bounded by KoAT2 or CoFloCo, and KoAT2 reports a linear bound for 96 of the 133, a quadratic one for 34 and a constant one for 3. Before quadratic terms were added to the checker the count was 103; a further 95 programs have a potential for every loop but no certificate (139 before). For the comparison of coverage we also ran the tool with every integer parameter, and every value assigned from an undefined function, restricted to $[-1000, 1000]$ by annotations added by a script. A bound found in this mode holds for those inputs only, so it is compared with the other tools only on which programs receive a bound. In this mode CrossCheck bounds 215 programs (171 before the search for a potential was repeated with more candidate invariants, and 207 before the analysis kept the facts true in both arms of a conditional expression and split a disequality in a loop guard into two strict inequalities); KoAT2 and CoFloCo, run as on TACLeBench, bound 288 and 264 programs for all inputs (305 by at least one of them). Of the 215 programs bounded by CrossCheck, 198 are also bounded by KoAT2 or CoFloCo. Eleven of the other 17 are bounded only because of the ranges: they assign a nondeterministic value to a variable inside the loop, so they have no bound in terms of their inputs, and the other tools are right to report none (one of them does not terminate for unbounded inputs). On a nest of four loops both tools timed out. The remaining five have simple linear potentials once the ranges of the variables are invariants, for instance $x + y + 1000$ for a loop that swaps and decrements two variables; the certificates of these five programs, with those potentials, are accepted by the checker. 107 programs are bounded by KoAT2 or CoFloCo and not by CrossCheck. In a sample of ten programs of this kind, drawn when there were 147, two were bounded once the analysis kept facts across \texttt{break} exits; in the other eight the missing ingredients are bounds that are not given by one linear potential (guards $x \ne y$, counters that are reset, loops whose direction changes) and relations carried from one loop to the next (the bound of a loop depends on a value computed by an earlier loop, or nested loops share a counter). A second sample of ten, drawn when there were 115, gives the same picture: three guards with a disequality or a disjunction, five nests that share counters and need amortisation, one relation carried out of an inner loop, and one conditional expression, now bounded. They correspond to the linear template and the dependence on invariants discussed in Section~\ref{sec:limitations}; the comparison measures their effect on programs written to exercise complexity analyzers. With certificates, 159 of the 215 entry functions bounded in the second mode are certified, all accepted by the checker: 97 with a summary affine in their parameters, valid for every input in the given ranges (for 74 of them KoAT2 also gives a linear bound), and 62 with a constant bound (17 of these had a summary before the amortised loop rules: the generator compares its candidates at the largest admissible inputs, where the amortised constant bound is about half the summary, so that a bound in the parameters is traded for a tighter one at the extreme inputs); the functions declared without a body are assumed contracts of cost $1$ (46 contract entries). No program exceeded the time limit of 120 seconds in this run. In this mode the summaries are obtained with the help of the ranges; the mode without ranges, above, is the one whose certified bounds hold for all inputs. The harness of the bounded mode changes what is measured (inputs in a range, nondeterministic values bounded too), which is why only coverage is compared in that mode.

\section{Related Work}
\label{sec:related}

The theoretical background (cost logics, ranking functions, closed forms, amortization) is discussed in \cite{crosscheckTheory}; this section is about tools.

\paragraph{Bound analyzers.}
Automatic bound analyzers derive symbolic bounds from ranking functions, counters and invariants \cite{gulwani2009speed,alias2010multi,sinn2014complexity,brockschmidt2016analyzing,lommen2022automatic}, from cost relations solved into closed forms \cite{albert2011closed,floresmontoya2014cofloco}, or from potential-based amortized analysis \cite{hoffmann2012multivariate,carbonneaux2015compositional}. Loopus \cite{sinn2014complexity} obtains amortized bounds for nested loops through lexicographic ranking functions, which the plugin does not attempt: its nested loops are bounded by products, and only its certificates amortise two triangular shapes. Bounding non-rectangular loop nests by summation is classical in WCET analysis \cite{healy2000supporting}, and summing the costs of successive iterations as an arithmetic sequence is the method of \cite{albert2011more}. Two of them, KoAT2 and CoFloCo, are compared with the plugin in Section~\ref{subsec:tacle}, together with Infer's cost analysis \cite{infercost} and LLVM's ScalarEvolution \cite{lattner2004llvm}; on programs written for complexity analysis they are clearly stronger. The plugin's place is a different one: numeric bounds of real-time code at compile time, in calling contexts, with certificates.

\paragraph{WCET flow analysis.}
Industrial WCET analyzers combine a flow analysis, which bounds loops and excludes infeasible paths, a micro-architectural analysis, and a path analysis, usually by implicit path enumeration (IPET) with integer linear programming \cite{wilhelm2008worst,li1995ilp}. oRange \cite{demichiel2008orange} computes loop bounds for this purpose by propagating values from the program's entry point; its bounds coincide with the plugin's in-context bounds wherever both exist (Section~\ref{subsec:tacle}). The plugin composes costs by the timing schema \cite{shaw1989reasoning,puschner1989calculating} rather than by IPET, and has no micro-architectural model; its loop bounds are exported as flow facts for such tools (Section~\ref{sec:implementation}, paragraph ``Exports'').

\paragraph{Verified and certified bounds.}
Blazy, Maroneze and Pichardie verified a loop-bound estimation in Coq \cite{blazy2013formal}, and Maroneze et al.\ built on it a formally verified WCET estimation, with loop bounds and IPET, integrated in the CompCert compiler \cite{maroneze2014formally}. Carbonneaux et al.\ obtained certified resource bounds for C programs \cite{carbonneaux2015compositional}. These works prove properties of the analysis or of its result inside a proof assistant (Coq), on the intermediate languages of CompCert (RTL for the loop bounds, Clight for the resource bounds). The plugin instead follows the pattern of proof-carrying code \cite{necula1997proof}: an unverified producer emits a certificate and a small checker, proved sound once \cite[Theorem~36]{crosscheckTheory}, validates it. The checker is simpler than a verified analyzer and does not depend on how certificates were found, but the translation from C++ to the certificate's program is outside the proof, where the CompCert-based tools are verified down to the compiler's representation. Validating each certificate against an independent description of the source, as translation validation does for compilers \cite{pnueli1998translation,necula2000translation}, removes the generator from the trusted base for the functions it covers, though not the translation of C into that description.

\paragraph{Compile-time real-time checks.}
Recent versions of Clang check, through function effect attributes, that functions declared non-blocking or non-allocating do not call functions that may block or allocate \cite{clangeffects}. Such checks exclude unbounded operations but do not bound loops or recursion; the plugin's real-time profile (Section~\ref{sec:implementation}) rejects allocation, exceptions and indirect calls, and adds the bounds.

\section{Limitations}
\label{sec:limitations}

\paragraph{Cost is not time.}
The bounds of this paper are bounds on a cost norm. With unit weights they count instructions; with weights $\omega$ they bound execution time only insofar as $\omega(a)$ is a valid worst-case duration of instruction $a$ in every context. On processors with caches, pipelines and branch prediction, instruction durations are context-dependent, and obtaining valid weights is the task of micro-architectural analysis \cite{wilhelm2008worst}. CrossCheck provides the loop-bound and path-cost components of a WCET analysis, not a complete WCET analyzer.

\paragraph{Precision and coverage of the plugin.}
Potentials are affine or logarithmic: a loop that divides a variable by a constant at every iteration, or multiplies it by a constant up to a constant limit, receives a logarithmic bound, certified when its entry bound is constant and it has no early exit, but binary search (whose potential is a difference of two variables) receives a linear bound, a lexicographic argument is found only when its inner bound is a constant, and loops that need several phases or a disjunction receive a linear bound or none; nested loops are bounded by products without amortization, except in certificates for a nest whose inner loop runs from the outer counter to a constant limit or from $0$ to the outer counter, with constant entry bounds; calling contexts are intervals of integer arguments, without relations between arguments, values of global variables or effects of callees on memory. Pointers, iterators and floating-point values are not tracked as loop counters. The certificate fragment excludes recursion, \texttt{do} loops with a jump or a \texttt{switch} in their body, \texttt{goto} and trusted annotations. Validation against the source covers a smaller fragment: no calls in tests, no call that the generator inlined (the source has the call), and the elimination of jumps in a \texttt{do} loop only when its body has no \texttt{break} or \texttt{continue} of its own; and a validated function is described down to its source only if the functions it calls are validated too. In the mode without ranges, certified bounds are at most quadratic in the parameters: a nest of three loops over parameters has no certificate, nor has a product loop whose potential decreases by more than one. The search for invariants re-analyses inner loops in every round of the refinement of the enclosing loops, so its cost grows with the depth of the nesting; each function has a search budget, counted in the size of the solver queries so that results do not depend on the machine, beyond which the optional searches stop and the refinement keeps only the invariants already proved inductive (at most two rounds). The results are then less precise, never unsound, and the report says so; on TACLeBench no bound changed with the default budget.

\paragraph{Evaluation.}
The prototype has been run on the small programs of Table~\ref{tab:plugin}, written for the purpose, on the 29 kernels of TACLeBench (Table~\ref{tab:tacle}), which are small programs, on its 25 sequential and application programs and on two programs of its parallel category derived from industrial systems (ROSACE and DEBIE), for which loop counts were not observed, and on the CMSIS-DSP library, with assumed input ranges and without observed counts or certificates. The comparison there is with the flow facts shipped with the benchmarks, which hold for the fixed inputs of the programs, with loop counts observed on those inputs, and with five existing analyzers run without annotations or tuning (Table~\ref{tab:tools}); among WCET flow-analysis tools only oRange was compared. On the C programs of the Termination and Complexity Competition, CrossCheck bounds far fewer programs than KoAT2 and CoFloCo, with input ranges supplied by a script or, without ranges, with a harness that makes the values read before the first loop parameters. The random programs of the tests against executions are small and come from a generator written by the same hand as the tool; they complement the benchmarks and do not replace an independent test suite. The ACSL export was not given to a deductive verifier, and the flow facts were not loaded into aiT (their use with OTAWA is described in \cite{crosscheckBinary}). Industrial code bases and the effect of annotations are future work.

\section*{Conclusions}

A compiler plugin can bound loops and recursion of real-time C and C++ functions at compile time, and, for a linear fragment, emit certificates that a checker proved sound in Lean validates: on the TACLeBench kernels, 161 of the 165 bounded functions are certified, and no bound of the plugin is below a loop count observed by running the programs. Its loop bounds in the calling contexts of the programs agree with those of oRange, a dedicated loop-bound analyzer for WCET analysis, on all but one of the loops that both bound (binary search, bounded linearly). On programs written for complexity analyzers it remains weaker than KoAT2 and CoFloCo: certified bounds are affine in the parameters, and potentials are affine or logarithmic.

Future work includes extending the certificate fragment (recursion, \texttt{do} loops with jumps, bounds of degree above two in the parameters) and the fragment validated against the source (static locals, calls in tests, calls inlined by the generator), amortized bounds for nested loops beyond the two triangular shapes, flow facts that OTAWA and aiT can place on optimised code without a separate loop matching, with a timing model of the processor, a Rust front-end on a stable compiler interface with a larger fragment, and an evaluation on larger programs.

\section*{Artifact}

The Clang plugin, the checker program, the example programs and the scripts that fetch the benchmarks and produce the tables of this paper are in release v1.0 of the repository \url{https://github.com/ldallolio/crosscheck}, archived at Zenodo, \href{https://doi.org/10.5281/zenodo.24425070}{doi:10.5281/zenodo.24425070}, and at Software Heritage, \href{https://archive.softwareheritage.org/swh:1:rel:70c77563037e6db2a55b48b3ad4963c86a67773f}{\nolinkurl{swh:1:rel:70c77563037e6db2a55b48b3ad4963c86a67773f}}.

\section*{Statement on the Use of AI}

Following the recommendations on the responsible release of AI-generated mathematics \cite{agmai2026responsible}, we document how AI was used in this work. In November 2026 an AI coding assistant (Anthropic's Claude, models Opus~5.5 and Fable~5.2, and Haiku~5.5 inside the component that reads web pages, used through the Claude Code tool) wrote the Clang plugin, its tests, its example programs, the certificate generator, the mini-C exporter, the checker program, the run with minimal annotations, the Rust front-end, the generator of random test programs, the interpreter of certificates and the mutation test, the ACSL export, ran the measurements and the comparisons with other tools of this paper, and drafted its text; the prompts and a summary of the changes are recorded in the file \lean{AI\_USAGE.md} of the repository. The assistant also compared the sentences that cite a paper with the copies of the cited works that it could obtain. I have read and verified the text and take full responsibility for the content.

Thanks are also due to Prof. E. Denti, Dr. B. Franceschiello and Prof.
A. Omicini for their valuable suggestions and discussions.

\bibliographystyle{unsrt}
\bibliography{crosscheck-tool}

\end{document}